\input harvmac 
\input epsf

\def\CC{{\cal C}}

\lref\gvconjecture{R. Gopakumar and C. Vafa, 
``On the Gauge Theory/Geometry Correspondence,''
{\tt hep-th/9811131}.}
\lref\fp{C. Faber and R. Pandharipande,
``Hodge Integrals and Gromov-Witten Theory,''
{\tt math.AG/9810173}.}
\lref\candelas{P. Candelas, X.C. De La Ossa, P. S. Green and
L. Parkes, ``A Pair of Calabi-Yau Manifolds as an Exactly Soluble 
Superconformal Theory,'' Nucl. Phys. B359 (1991) 21.}
\lref\bcovone{M. Bershadsky, S. Cecotti,
H. Ooguri and C. Vafa, ``Holomorphic Anomalies
in Topological Field Theories,'' Nucl. Phys. B405 (1993) 279;
{\tt hep-th/9302103}.}
\lref\bcovtwo{M. Bershadsky, S. Cecotti,
H. Ooguri and C. Vafa, ``Kodaira-Spencer Theory
of Gravity and Exact Results for Quantum 
String Theory,'' Commun. Math. Phys. 165 (1994) 311; 
{\tt hep-th/9309140}.}
\lref\wittenopen{E. Witten, 
``Chern-Simons Gauge Theory as a String Theory,''
{\tt hep-th/9207094}.}
\lref\wittencs{E. Witten,
``Quantum Field Theory and the Jones Polynomial,''
 Commun. Math. Phys. 121 (1989) 351.}

\noblackbox

\Title{\vbox{\baselineskip12pt
\hbox{\tt CALT-68-2251, CITUSC/99-008}
\hbox{\tt HUTP-99/A070, LBNL-44695}
\hbox{\tt UCB-PTH-99/54}
\hbox{\tt hep-th/9912123}}}%
{\vbox{\centerline{Knot Invariants and Topological Strings}}}
\bigskip
\centerline{Hirosi Ooguri$^\dagger$\foot{On leave of absence
from University of California, Berkeley.} and Cumrun Vafa$^*$}
\bigskip
\smallskip
\centerline{$^\dagger$
Caltech - USC Center for Theoretical Physics, Mail Stop 452-48}
\centerline{California Institute of Technology, Pasadena, CA 91125}
\medskip
\centerline{$^*$ Jefferson Laboratory of Physics,
Harvard University, Cambridge, MA 02138}

\vskip 0.5cm
\centerline{\tt  hooguri@lbl.gov, vafa@string.harvard.edu}

\vskip .3in
We find further evidence for the conjecture
relating large $N$ Chern-Simons theory on $S^3$
with topological string on the resolved conifold geometry
by showing that the Wilson loop observable of a simple knot
on $S^3$ (for any representation) agrees to all orders in $N$
with the corresponding quantity on the topological string side.
For a general knot, we find a reformulation of the knot invariant
in terms of new integral invariants, which capture the spectrum (and spin)
of M2 branes ending on M5 branes embedded in the resolved
conifold geometry.  We also find an intriguing link between knot invariants
and superpotential terms generated by worldsheet
instantons in $N=1$ supersymmetric theories
in 4 dimensions.

\Date{December 1999}

\newsec{Introduction}
Large $N$ gauge theories have been conjectured
by `t Hooft to be related to string theories.
A particularly simple example of gauge theories
is the Chern-Simons theory, solved by Witten.  It is thus
natural to ask about the large $N$ limit of Chern-Simons
theory and look for an appropriate stringy description.
Some aspects of large $N$ limit of Chern-Simons theory were studied some
time ago
in \ref\peri{V. Periwal, ``Topological Closed-string Interpretation
of Chern-Simons Theory,'' Phys. Rev. Lett. 71 (1993) 1295.},
\ref\doch{
M.R. Douglas, ``Chern-Simons-Witten Theory as a Topological
Fermi Liquid,'' hep-th/9403119.}.

It was conjectured recently \gvconjecture\ that at least
for some manifolds (including $S^3$) the large $N$ limit does give
rise to a topological string theory on a particular
Calabi-Yau background.  This conjecture
was checked at the level of the partition function on both
sides;  The Chern-Simons answer was already well known, and
the topological string partition function was recently
computed in two different ways (one by mathematicians, and one by using 
physical reasoning about the structure of BPS states).

It is natural to extend the conjecture to the observables of
Chern-Simons theory, which are Wilson loop operators. Namely
we should consider product of Wilson loop observables
for any choice of representation on each knot. We show
how this question can be formulated in the present context
and explicitly check the map for the case of the simple circle in
$S^3$ (``unknot'').  The computation on the Chern-Simons side
is well known.  On the topological string side, we end up
with topological string amplitudes on Riemann surfaces with boundaries.
Mathematically these have not been studied, however 
by connecting the partition function of topological strings to target space
quantities we compute them in terms of spectrum of M2 branes ending
on M5 branes embedded in the Calabi-Yau threefold.  The target
space interpretation is also related to generation of $N=1$ superpotential
terms in four dimensions (which we relate it analogously
to the spectrum of domain walls).

For a general knot finding the explicit form of the
M5 brane embedded in the Calabi-Yau is not trivial, though
physically we argue it should be possible. In this
way we reformulate knot invariants in terms of new invariants capturing
 the spectrum of M2 branes ending on M5 branes.

The organization of this paper is as follows: In section 2 we review
the large $N$ conjecture for Chern-Simons theory.  In section 3 we show
how the Wilson Loop observable for arbitrary knot and representation
can be formulated in this set up, and apply the gauge theory/geometry
correspondence for the case of the simple knot.  In section 4 we
show how the results anticipated from the Wilson loop observables
can be directly obtained using the spectrum
of M2 branes ending on M5 branes (or D2 branes ending on D4 branes).
We also point out connections with generation of superpotential
terms with theories with 4 supercharges.  In section 5 we present
some concluding remarks and suggestions for future work.

\newsec{The Large $N$ conjecture for Chern-Simons Theory }

In this section we review the conjecture of \gvconjecture\ 
which relates large $N$ limit of $SU(N)$ Chern-Simons
gauge theory on $S^3$ to a particular topological string
amplitude.  The motivation for the conjecture
was that, in the context of topological strings
of the $A$-type on Calabi-Yau threefolds, there are D-branes with 
three-dimensional worldvolume which support the Chern-Simons gauge theory
\wittenopen .  So it is natural to expect that at least in some
cases, by putting many branes on some cycles and taking the large $N$ 
limit,
we end up with a topological string on some deformed Calabi-Yau, 
but without branes.
This is what was found to be the case in \gvconjecture , which we will
now review.

\subsec{The statement of the conjecture}

The conjecture in \gvconjecture\ 
states that
the Chern-Simons gauge theory on $S^3$ with gauge
group $SU(N)$ and level $k$ is equivalent to
the closed topological string theory of $A$-type
on the $S^2$ blown up conifold geometry with 
\eqn\parameters{\lambda = {2\pi \over k+N}, ~~~~ 
t={2\pi i N \over k+N},}
where $\lambda$ is the string coupling constant
and $t$ is the K\"ahler modulus of the blown-up $S^2$. 
The coupling constant $g_{CS}$ of the Chern-Simons theory, 
after taking into account the finite renormalization, is
related to $\lambda$ as $\lambda=g_{CS}^2$. Therefore the K\"ahler moduli $t$
given by \parameters\ is $i$ times 
the 't Hooft coupling $g_{CS}^2 N$ of the 
Chern-Simons theory. The geometric motivation of the
conjecture is based on starting with the topological strings on conifold
geometry $T^*S^3$ and
putting many branes on $S^3$, for which we get a large $N$
limit of Chern-Simons on $S^3$ supported on the brane.
The conjecture states that in the large
$N$ limit the branes disappear but lead instead to the
resolution of the conifold geometry where an $S^2$ has
blown up. 
In fact this conjecture parallels the motivation
for the AdS/CFT
correspondence conjecture:  As noted in \gvconjecture, 
since the open topological string theory couples to closed
topological string theory through a gravitational
Chern-Simons action \wittenopen , putting 3-branes
on $S^3$ deforms the gravitational background
so as to produce a blown up $S^2$.  In fact
the volume of the $S^2$ was computed in this way.

The conjecture has been checked as follows: Start with 
the vacuum amplitude $Z(S^3)$ of the 
Chern-Simons gauge theory on $S^3$ (with the normalization
$Z(S^2 \times S^1) = 1$);
\eqn\cspartition{Z(S^3) = {e^{i{\pi \over 8}(N-1)N} \over (k+N)^{N/2}}
\sqrt{{k+N \over N}} \prod_{s=1}^{N-1} \left[2 \sin\left({s\pi\over k+N}
\right) \right]^{N-s}.}
The large-$N$ expansion of $\log Z(S^3)$ is given by
\eqn\largenpartition{
Z(S^3) = \exp\left[ - \sum_{g=0}^\infty \lambda^{2g-2} F_g(t) \right],}
where $\lambda$ and $t$ as in \parameters ,
\eqn\zeroandone{
\eqalign{ F_0 & = -\zeta(3) + {i\pi^2\over 6} t -i\left( m + {1\over 4}\right)
\pi t^2 + {i \over 12}t^3 + \sum_{n=1}^\infty n^{-3} e^{-nt} \cr
F_1 & = {1 \over 24} t + {1 \over 12} \log \left( 1-e^{-t} \right) ,\cr}}
with $m$ being some integer, and for $g \geq 2$, 
\eqn\twoorlarger{
F_g = {(-1)^{g-1} \over 2g(2g-2)} B_g \left[ 
{(-1)^{g-1} \over (2\pi)^{2g-2}} 2\zeta(2g-2)
        - {1 \over (2g-3)!} \sum_{n=1}^\infty n^{2g-3} e^{-nt}
\right].}
Here $B_g$ is the Bernoulli number,
which is related to the Euler characteristic of the moduli space
${\cal M}_g$ of genus-$g$ Riemann surfaces as
\eqn\euler{\chi_g ={(-1)^{g-1} \over 2g(2g-2)} B_g.}
By using this and the formula for the Chern-class
of the Hodge bundle over the moduli space 
\eqn\chern{  \int_{{\cal M}_g} c_{g-1}^3 = 
{(-1)^{g-1} \over (2\pi)^{2g-2}} 2\zeta(2g-2) \chi_g,}
which was derived in \fp, one can rewrite \twoorlarger\ as
\eqn\closedinstantons{F_g = \int_{{\cal M}_g} c_{g-1}^3
-  {\chi_g \over (2g-3)!} \sum_{n=1}^\infty n^{2g-3} e^{-nt}.}
It turns out that the expressions \zeroandone\ and \closedinstantons\
for $F_g$ are exactly those of the $g$-loop topological
string amplitude on the resolved conifold. The constant map from
the worldsheet to the target space gives rise to the term
${i \over 12} t^3 $ in $F_0$ \candelas\foot{
The coefficients of $1$ and $t$ in $F_0$ have analogous interpretations
\ref\yauetal{S. Hosono, A. Klemm, S. Theisen and S.-T. Yau,
``Mirror Symmetry, Mirror Map and Applications to Complete
Intersection Calabi-Yau Spaces,'' Nucl. Phys. B433 (1995) 501;
{\tt hep-th/9406055}.}, and they also agree with the Chern-Simons 
prediction \gvconjecture .}, and ${1 \over 24} t$ in $F_1$
\bcovone, and $ \int_{{\cal M}_g} c_{g-1}^3$ in $F_g$ with $g \geq 2$
\bcovtwo . Regarding worldsheet instantons, since 
the only non-trivial 2-cycle in the target space is the 
blown-up $S^2$, their contributions are from multi-coverings of 
the Riemann surface onto the $S^2$. For $g=0,1$ and $2$,
the expressions in the instanton terms in \zeroandone\ and
\twoorlarger\ agree with the results of \candelas, \bcovone, and 
\bcovtwo\ respectively. More recently, the contribution
of of the worldsheet instantons are evaluated for all $g$
in \fp , in complete agreement with \closedinstantons .
These expressions can also be derived, as was done in \ref\gopva{
R. Gopakumar and C. Vafa, ``M-Theory and Topological Strings I, II,''
{\tt hep-th/9809197, 9812127}.}, from the target space view point by
identifying what the topological strings compute in Type IIA
compactification on the corresponding Calabi-Yau threefold.  
It turned out that the full structure of $F_g$ is encoded 
in terms of the spectrum of wrapped D$2$ branes
on the Calabi-Yau. This will be reviewed later in this paper.

In this paper, we provide further evidence for the conjecture.
We will show that the Wilson loop expectation value of the 
Chern-Simons theory also has a natural interpretation in
terms of the topological string on the resolved conifold geometry. 

\subsec{Conifold transition}

As noted above, the geometric insight that led to the conjecture 
is the fact that one can view the Chern-Simons
theory as the open topological string theory. Consider the
cotangent space $T^* S^3$ of $S^3$ as the target space of
the topological string. It was shown in \wittenopen\
that, if we wrap $N$ D-brane on the base $S^3$
of the cotangent space, the open topological string theory
on the D-brane is equivalent to the Chern-Simons theory
with the gauge group $SU(N)$. The cotangent space has
the canonical symplectic form
\eqn\symplectic{ \omega = \sum_{i=1}^3 dq^i \wedge dp_i,
~~~(q \in S^3,~ p \in T_q S^3),}
and the base $S^3$ is a Lagrangian submanifold. 
Therefore the open string on the D-brane allows the topological 
twist of the $A$-type. 

At this point, it would be useful to review basic facts about
the conifold transition. 
The space $T^* S^3$ can also be regarded as a deformed conifold geometry,
\eqn\deformedconifold{ \sum_{\mu = 1}^4 y_\mu^2 = a^2,~~~~(y \in C^4)}
where, without loss of generality, we assume the deformation
parameter $a$ to be real. To see that \deformedconifold\ is indeed $T^* S^3$, 
we can set $y_\mu = x^\mu + i p_\mu$
and rewrite the equation as
\eqn\realform{ (x^\mu)^2 - (p_\mu)^2 = a^2,~~~
x^\mu p_\mu = 0. }
The first equation suggests that the base $S^3$ of radius $a$ is
located at $p_\mu = 0$ and the second equation shows that
$p_\mu$ are coordinates on the cotangent space at $x \in S^3$. 
As $a \rightarrow 0$, the $S^3$ shrinks to a point
and a singularity appears. It is known as the conifold singularity. 

In addition to the deformation by $a$, the conifold singularity
\eqn\conifold{\sum_\mu y_\mu^2 = 0}
can be smoothened out by the blow-up. It is described as follows. 
By introducing two pair of complex coordinates $(u, \tilde u)$
and $(v, \tilde v)$ by
\eqn\newcoords{\eqalign{ &u = y_1 + i y_2, ~~~ \tilde u = y_3 - i y_4 \cr
 & v = y_3 + i y_4, ~~~ \tilde v = y_1 - i y_2, \cr}}
the equation \conifold\ can be written as
\eqn\conifold{ u \tilde v + v \tilde u = 0.}
This means that there is some $z$ such that
\eqn\transition{u = z \tilde u, ~~~ v = -z \tilde v.} 
If we view $z$ as a complex coordinate on $S^2$ (as we should since
$\tilde u$ can be $0$ and we need to add $z=\infty$), 
one can interpret \transition\ as defining the bundle
${\cal O}(-1) + {\cal O}(-1)$ over $S^2$ where
$u$ and $v$ are coordinates on the fibers. With respect to
the original symplectic form \symplectic , the volume of
$S^2$ is zero, which is another way to see that the
conifold geometry is singular. We can remove the singularity 
by blowing up the $S^2$; this process is called the small resolution
(as opposed to the deformation of complex structure $a$
in the previous paragraph). See
Figure 1. The conifold singularity \conifold\ 
can be either deformed to the total space of $T^* S^3$ or 
resolved to the total space of ${\cal O}(-1) + {\cal O}(-1)$ 
over $S^2$. The transition from one to the other is called 
the conifold transition.

The conjecture in \gvconjecture\  states that
the open topological string theory on the $N$ D-branes on
$S^3$ of the deformed conifold is equivalent to the closed topological
string theory (without D-branes) on the resolved conifold 
with $t = 2\pi iN/(k+N)$.

\bigskip

\centerline{\epsfxsize 3.9truein \epsfysize 3.3truein\epsfbox{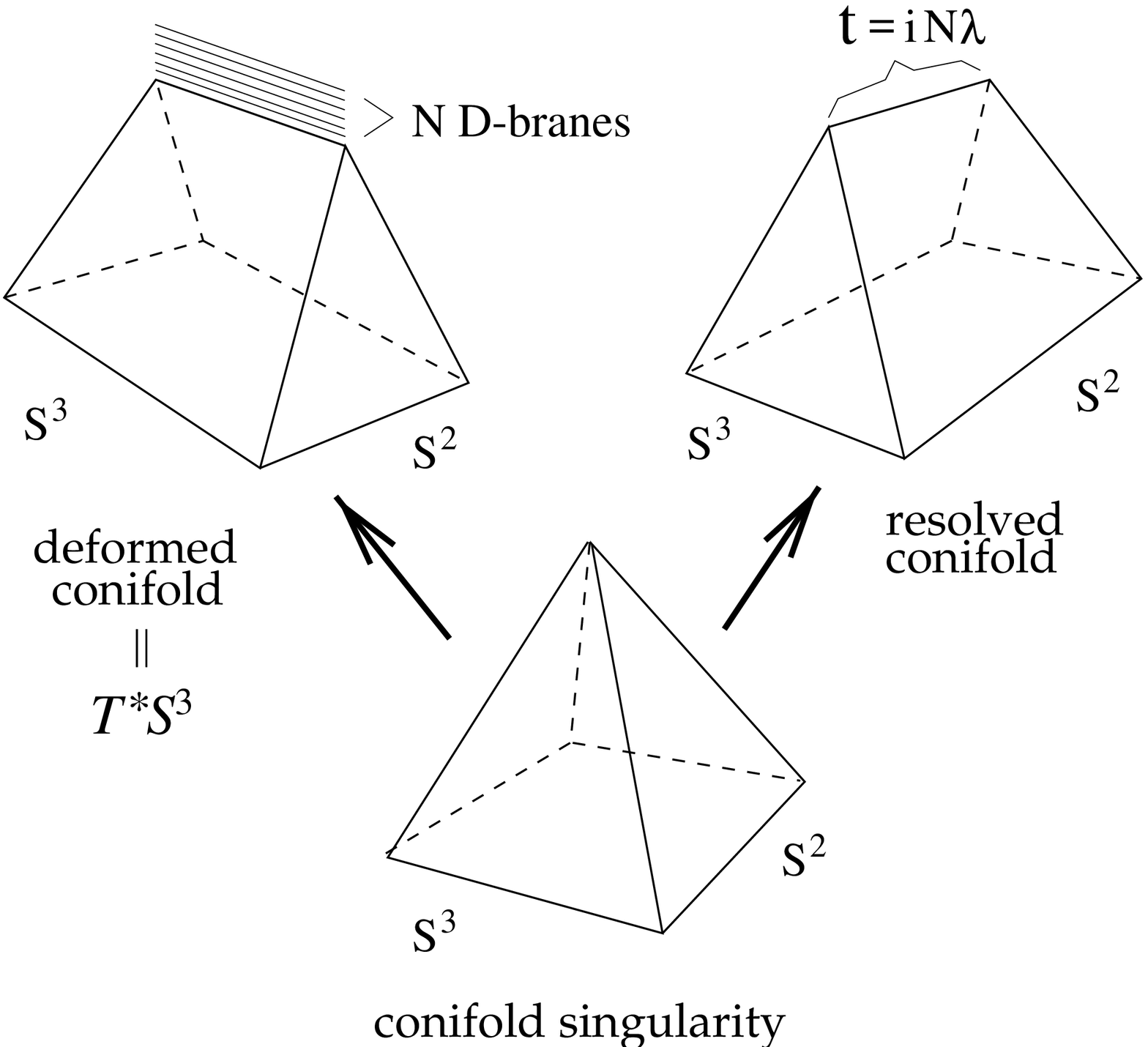}}
\noindent{\ninepoint\sl \baselineskip=8pt {\bf Figure 1}: {\sl
The conifold singularity can be either deformed to 
$T^*S^3$ or resolved by the $S^2$ blow-up.
The conjecture in \gvconjecture\
 states that the open topological string theory on the $N$ D-branes on
$S^3$ of the deformed conifold is equivalent to the closed topological
string theory on the resolved conifold geometry.}}

\bigskip

\newsec{Wilson loop}

\subsec{Definition}

The basic observables of the Chern-Simons gauge theory
are the Wilson loops. For each loop $q(s) \in S^3$ ($0 \leq s < 2\pi$), 
we can define  a generating function of Wilson loops of various 
representations of $SU(N)$ in the following way. As shown in \wittenopen ,
the Chern-Simons theory is the open topological string theory
on $N$ D-branes wrapping the base $S^3$ of $T^* S^3$. We can
probe the dynamics on these D-branes by introducing another
set of D-branes. First we define a Lagrangian 3-cycle associated
to the knot $q(s) \in S^3$ as follows \foot{We thank C. Taubes
for discussion on these Lagrangian cycles.}. At each point $q(s)$
on the loop, we consider 2-dimensional subspace of $T_q^* S^3$
orthogonal to $dq/ds$. By going around the loop, we can define
the 3-cycle,
\eqn\lagrangianforloop{
  \CC = \{~ (q(s), p)~ |~~ p_i {dq^i\over ds} = 0, ~~0 \leq s < 2\pi ~\}.}
The topology of ${\cal C}$ is $R^2 \times S^1$.
The symplectic form $\omega$ vanishes on ${\cal C}$, 
so it is a Lagrangian
submanifold\foot{The cycle ${\cal C}$ defined here is Lagrangian 
but is not necessarily special Lagrangian.
In order for to make the topological $A$-twist, it is sufficient
that $\omega$ vanishes on $\CC$.}.
The 3-cycle intersects with the base $S^3$ along the loop $q(s)$.
See Figure 2.

\bigskip

\centerline{\epsfxsize 2.5truein \epsfysize 2.3truein\epsfbox{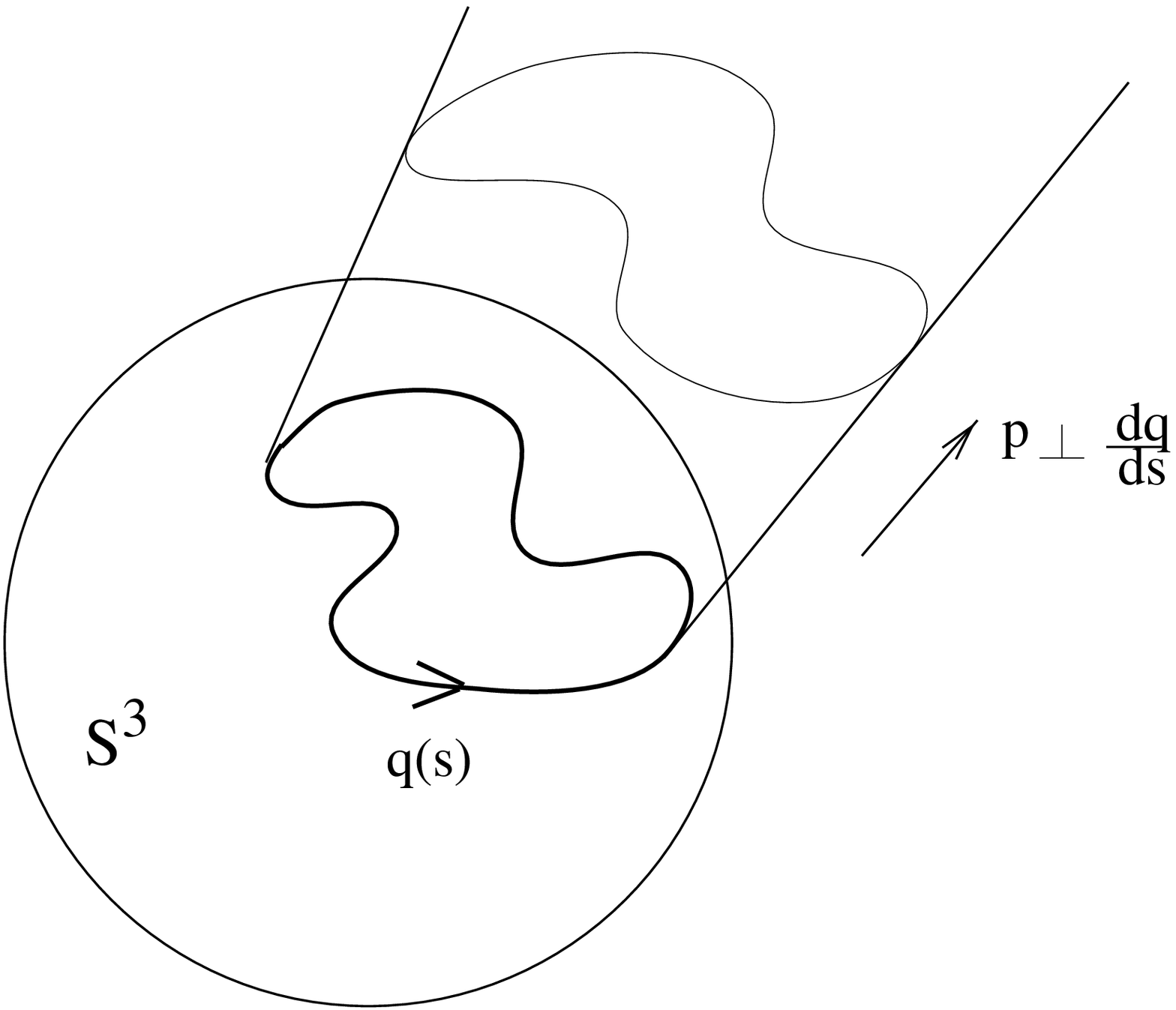}}
\noindent{\ninepoint\sl \baselineskip=8pt {\bf Figure 2}: {\sl
For each loop $q(s) \in S^3$, one can define a unique Lagrangian
3-cycle which extends in the cotangent direction and intersects
$S^3$ on the loop $q(s)$.}}

\bigskip

Now let us wrap $M$ D-branes on $\CC$. We then have
 the $SU(M)$ Chern-Simons theory on $\CC$ as well as
the $SU(N)$ Chern-Simons theory on $S^3$.
In addition, we also have a new sector of open string with one end 
on $\CC$ and the other on $S^3$. One can easily quantize the topological
string in this sector and obtain a complex scalar field living
on the intersection, namely the loop $q(s)$, which transforms
according to the bi-fundamental of $SU(N)\otimes SU(M)$.
To see that there is one complex scalar field of this type, we note
that, in the relevant open string sector $i.e.$ the Ramond sector, 
there are two states one with $N=2$ $U(1)$ charge $-1/2$ (a scalar) 
and the other with $+1/2$ (a 1-form).  The physical
states of the topological string come from the sector with $U(1)$
charge $-1/2$, and that turns out to correspond to
the scalar living on the loop $q(s)$.
 The action for the scalar field is Gaussian,
and integrating it out gives the determinant,
\eqn\determinant{ Z = 
\exp\left[- \log \det 
\left( {d\over ds} + (A_i - \tilde{A}_i) {dq^i\over ds} \right) \right],}
where $A$ and $\tilde A$ are the Chern-Simons gauge fields
on $S^3$ and $\CC$ respectively. We can evaluate the determinant
by diagonalizing $A$ and $\tilde A$ (this is allowed since 
we are dealing with the one-dimensional problem along the
intersection). By using the formula
\eqn\detformula{ \log \det\left[ {d \over ds} + i \theta \right]
  = \sum_{n=-\infty}^\infty \log (n+\theta)  
  = \log \sin(\pi \theta) + {\rm const},} 
we find
\eqn\detresult{\eqalign{
 Z(U,V) & = \exp\left[ - \tr \log \left( U^{-1/2} \otimes V^{1/2} 
- U^{1/2} \otimes V^{-1/2} \right) \right] \cr
& = \exp\left[ - \tr  \log \left( 1 - U \otimes V^{-1} \right) \right]\cr
& = \exp\left[ \sum_{n=1}^\infty {1 \over n} \tr U^n
\tr V^{-n} \right], \cr
} }
where $U$ and $V$ are path-ordered exponentials of the gauge
fields along the loop, 
$$U = P\exp\oint A \in SU(N),~~~V = P\exp\oint \tilde A \in SU(M), $$
and we used $\det U = \det V = 1$.  

We are interested in taking $N$ to infinity for a fixed $M$.
We view the $M$ branes on $\CC$ as a probe.  In this context,
integrating out the gauge fields $A$ will leave us with an
effective theory on the probe brane, which is an $SU(M)$ Chern-Simons
on $\CC$ plus some corrections.  Let us define
\eqn\givea{
\eqalign{ {\rm exp}(-F(t,V)) & 
={1 \over  
\int [DA] \exp(- S_{CS}(A;S^3))}
 \int [DA] \exp\left[ - S_{CS}(A;S^3) + \sum_{n=1}^\infty {1 \over n} \tr U^n
\tr V^{-n} \right] \cr
&=\langle Z(U,V) \rangle_{S^3}\cr}}
which can be viewed as the generating functional
for all the observables of the Chern-Simons
gauge theory on $S^3$ associated to the unknot.  Then we obtain
an effective theory on the $\CC$ brane, which is the deformation
of the Chern-Simons theory as
\eqn\givv{S=S_{CS}(\tilde {A};\CC)+F(t,V).}
Here $S_{CS}(\tilde A; \CC)$ deonotes the Chern-Simons
action on $\CC$.

\subsec{Evaluation}

Let us evaluate $\langle Z(U,V) \rangle_{S^3}$ explicitly when
the loop $q(s)$ is the unknot, $i.e.$ a simple
circle in $S^3$ which represents a trivial
knot. In this case it is known
\wittencs\ that, for an admissible representation ${\cal R}_j$ of $SU(N)$,
the Wilson loop expectation value is 
\eqn\sformula{\langle \Tr_{{\cal R}_j} U \rangle_{S^3} = {S_{0j} \over S_{00}},}
where $S_{ij}$ is the modular transformation matrix for 
the characters of the $SU(N)$ current algebra at level $k$. 
If we know $\langle \Tr_{{\cal R}_j} U \rangle$ for all the representations,
we can compute any product of traces of $U^n$ in the fundamental
representation by using the Frobenius relation,
\eqn\frobenius{ \tr U^{n_1} \cdots \tr U^{n_h}
  = \sum_Y \chi(Y;, n_1, \dots , n_h) \Tr_{R(Y)} U, }
where $Y$ is the Young tableau with $n=n_1 + \cdots + n_h$
boxes, $R(Y)$ is the corresponding representation of $SU(N)$,
and $\chi(Y; n_1, \cdots, n_h)$ is the character of the
representation of the permutation group ${\cal S}_n$
corresponding to $Y$, evaluated for the permutation 
with cycles of sizes $n_1, \cdots, n_h$ (for example, see
\ref\gt{D.J. Gross and W. Taylor, ``Twists and Wilson Loops in
the String Theory of Two-Dimensional QCD,'' Nucl. Phys. B403 (1993) 395;
{\tt hep-th/9303026}.}). To actually evaluate 
$\langle \tr U^{n_1} \cdots \tr U^{n_h} \rangle_{S^3}$, we use
the following trick. We first note that $S_{0j}$ is given by  
\eqn\soj{ S_{0j} = \sum_{w \in {\cal W}} \epsilon(w)
  \exp \left[ -{2\pi i  
             \over k+N}(w(\rho), \lambda_j + \rho) \right]  ,   }
where ${\cal W}$ is the finite Weyl group of $SU(N)$, 
$\epsilon(w) = \pm 1$
is the parity of the element $w\in {\cal W}$, 
$\lambda_j$ is the weight vector for the representation $j$, and
$\rho$ is a half of the sum of positive roots. Therefore
$S_{0j}/S_{00}$ takes the form of the character of the
finite dimensional group $SU(N)$, namely
\eqn\character{
\langle \Tr_{{\cal R}_j} U \rangle_{S^3} = 
\Tr_{{\cal R}_j} U_0 ,}
where $U_0$ is a fixed element of $SU(N)$ which,
in the fundamental representation, takes the form
\eqn\cnumbermatrix{ U_0 = \left( \matrix{  
e^{\pi i (N-1)\over k+N} & 0 & 0& \cdots & 0 \cr
 0 &  e^{(N-3) \pi i \over k+N} & 0 & \cdots & 0 \cr
 0 & 0 & \cdots & \cdots & 0 \cr
 0 & 0 &  0 & \cdots  & e^{ \pi i(1-N) \over k+N} \cr}\right) .} 
Since $U_0$ in \character\ is the same for any ${\cal R}_j$,
to evaluate correlation functions of
the Wilson loops, we can simply replace $U$ by the $c$-number
matrix $U_0$ as
\eqn\productproduct{
 \langle \tr U^{n_1} \tr U^{n_2} \cdots \tr U^{n_h}
\rangle = \tr U_0^{n_1} \tr U_0^{n_2} \cdots \tr U^{n_h}_0 ,}
and
\eqn\loopevaluation{
 \tr U_0^n = 
{\sin\left( { nN\pi \over k+N}\right) \over \sin\left(
{n \pi \over k + N} \right)} = -i
 {e^{nt/2} - e^{-nt/2} \over 2\sin
\left(n \lambda/2 \right)}, }
where $\lambda$ and $t$ are as defined in \parameters . 
Substituting this back into \detresult, we find
\eqn\csprediction{
\langle Z(U, V) \rangle_{S^3}
= \exp \left[ -i\sum_{n=1}^\infty  
 {e^{nt/2} - e^{-nt/2} \over 2n \sin
\left(n \lambda / 2 \right)}  \tr V^{-n} \right].}
As we will see, this is exactly the form 
we expect for the topological string on the resolved
conifold geometry.

\subsec{Conifold transition of the Wilson loop}

In the case of the unknot, it is straightforward to
identify the effect of the conifold transition on the
Lagrangian submanifold $\CC$. Let us start with $T^* S^3$
expressed as \deformedconifold ,  
and consider the following anti-holomorphic involution on it.
\eqn\ztwo{ y_{1,2} = \bar{y}_{1,2},~~y_{3,4} = - \bar{y}_{3,4}.}
Since the symplectic form $\omega$ changes its sign under
the involution, its fixed point set is automatically a
Lagrangian submanifold of $T^* S^3$. This will be our $\CC$. 
If we write $y_\mu = x^\mu + i p_\mu$, the invariant locus
of the action \ztwo\ is
\eqn\fixed{ p_{1,2} = 0,~~ x^{3,4} = 0 }
and the equation \deformedconifold\ becomes
\eqn\fixedmore{(x^1)^2 + (x^2)^2 = a^2 + (p_3)^2 + (p_4)^2.}  
Therefore $\CC$ intersects with $S^3$ along
the equator of $S^3$, $i.e.$ the loop $q(s)$ is the unknot. The loop
$q(s)$ in this case is identified with 
$$q(s):\qquad  (x^1)^2+(x^2)^2=a^2, \quad x^3=x^4=0.$$

To define $\CC$ after the conifold
transition, we continue to identify it with the invariant
locus of the anti-holomorphic
involution.  To describe this explicitly let us
 use the coordinates $(u,v,z)$ or 
$(\tilde u, \tilde v, z^{-1})$ defined by \newcoords\ 
and \transition . In these coordinates,
the $Z_2$ invariant set $\CC$ is characterized by
\eqn\aftertransition{ u = \bar{\tilde v}, ~~v = -\bar{\tilde u},}
and the conifold equation \conifold\ restricted on $\CC$ becomes
\eqn\samesize{ u \bar u = v \bar v. }
The complex coordinate on the base $S^2$ defined by \transition\
is 
\eqn\phase{ z =- {u \over \bar{v}}.}
Because of \samesize, $z$ is pure phase.
Therefore one may view that $\CC$ is a line bundle
over the equator $|z| = 1$ of $S^2$ (the fiber being
the subspace of ${\cal O}(-1)+{\cal O}(-1)$ given by
$u + z\bar{v} = 0$).
In particular, $\CC$ intersects with the base $S^2$
along $|z|=1$. See Figure 3. Since the intersection is one-dimensional,
$\CC$ remains a Lagrangian submanifold 
even after the $S^2$ is blown up and the symplectic
form $\omega$ is modified.

\bigskip

\centerline{\epsfxsize 2.3truein \epsfysize 2.80truein\epsfbox{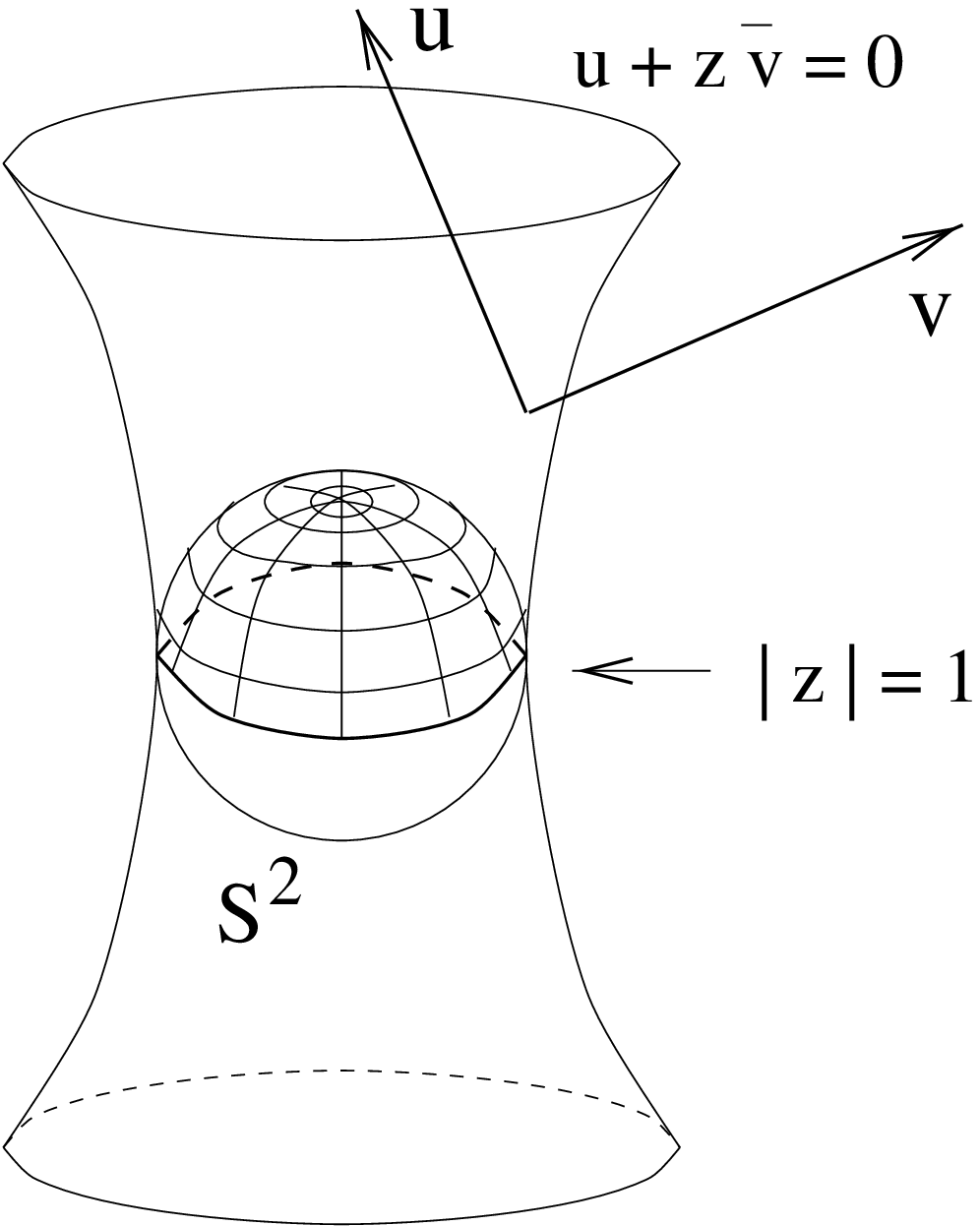}}
\noindent{\ninepoint\sl \baselineskip=8pt {\bf Figure 3}: {\sl
After the conifold transition, the Lagrangian 3-cycle touches
the base $S^2$ along the equator $|z|=1$ and extends in the fiber
directions following the constraint $u + z\bar{v} = 0$.
The worldsheet instanton can either wrap the northern
hemisphere, as shown in the figure, ending on the equator,
or wrap the southern hemisphere.}}

\bigskip

According to the conjecture of \gvconjecture ,
topological string with $N$ D-branes wrapping on the base 
$S^3$ of $T^* S^3$  
is equivalent to topological string on the resolved
conifold without D-branes. Here we are adding $M$ D-branes
on $\CC$ on one side, and have
traced it over to the other side.  On the $T^* S^3$ side, the effective
theory on the $M$ probe branes was the Chern-Simons action plus
some corrections \givv .  So the conjecture gives the falsifiable
prediction that, after the conifold transition, we should also
see the effective theory on the brane to be a deformed
version of the Chern-Simons theory \givv.  Indeed it has been shown
in \wittenopen\ that when there are holomorphic maps
from Riemann surfaces with boundaries  to the target space,
with boundaries lying on the D-brane, the Chern-Simons
action gets deformed.  In the original geometry of $T^* S^3$,
there are no such maps. However we got the deformation by integrating the
gauge theory on $S^3$ and the scalar field living on the knot.
At large $N$, we have made a transition to a new geometry
without any other sectors. But now, there {\it are}
non-trivial holomorphic maps that can end on $\CC$!
Since $\CC$ 
intersects with the base $S^2$ of the resolved conifold along 
$|z|=1$, there are holomorphic maps from Riemann surface
with a boundary with the image having the topology of disc.
 It is shown in \wittenopen\ that the effective
theory on $\CC$ should now be of the form
as predicted in \givv\ with $F(t, V)$ given by
\eqn\loopaftertransition{
  F(t,V)= \sum_{g=0}^\infty \sum_{h=1}^\infty 
\sum_{n_1,\cdots,n_h}^\infty
   \lambda^{2g-2+h} F_{g;n_1,\cdots,n_h}(t) 
 \tr V^{n_1} \cdots \tr V^{n_h} .}    
Here $F_{g;n_1,\cdots,n_h}$ is the topological string
amplitude on a genus-$g$ surface with $h$ boundaries. 
The factors $\tr V^{n_i}$
are picked up by the boundary of the worldsheet, which
wraps $|n_i|$-times the equator of the $S^2$ either clockwise 
$n_i > 0$ or counterclockwise $n_i < 0$ depending on whether 
the worldsheet is mapped to the upper or the lower hemisphere. 

To see that the Chern-Simons computation \csprediction\
agrees with this expectation, we note that, in the topological 
string computation, amplitudes are assumed to be analytic 
in $t$. By performing the analytic 
continuation\foot{By the analytic continuation, we can replace 
$\sum_{n=1}^\infty n^{2a} \tr V^n e^{nt/2}$
by $- \sum_{n=1}^\infty n^{2a} \tr V^{-n} e^{-nt/2} + {\rm const}$ for
any integer $a$.}, we can 
rewrite $\langle Z(U,V) \rangle$ as 
\eqn\cspredict{
\langle Z(U, V) \rangle_{S^3}
= \exp \left[ i\sum_{n=1}^\infty  
 {\tr V^n + \tr V^{-n}  \over 2n \sin
\left(n \lambda/ 2 \right)} e^{-nt/2}\right].}
This agrees with the general form \loopaftertransition\
expected for the topological string amplitude. 
We could make a more quantitative comparison by counting
holomorphic maps. 
There are only two basic holomorphic maps (with
the image being a disc) with
boundaries on $\CC$, which are
the upper and the lower hemispheres of the $S^2$, together
with their multicoverings, and with the higher genus coverings
of them (see Figure 3).
In particular, the comparison of \loopaftertransition\
and \cspredict\ suggests that all the relevant instantons have 
one boundary ending on $\CC$, wrapping the equator of $S^2$
either clockwise or counterclockwise. It would be interesting
to verify the prediction of the Chern-Simons computation
\cspredict\ explicitly using the worldsheet instanton calculus
extending the results from the closed string case to open strings. 
In this paper, we will take an alternative route, by giving
the target space interpretation of $F(t,V)$
and evaluate it explicitly by the Schwinger-type computation,
similar to what was done in the closed string case in
\gopva .
We will find that the prediction \cspredict\ is precisely reproduced
in this way. 

\newsec{Target Space Interpretation}

Topological strings are useful in computing superpotential-type 
terms in the context of superstring compactification
on Calabi-Yau manifold.  In particular it is known
\ref\bcovtwo, \ref\naret{I. Antoniadis, E. Gava, K. S. Narain
and T. R. Taylor, ``Topological Amplitudes in Superstring Theory,''
Nucl. Phys. B413 (1994) 162; {\tt hep-th/9307158}.}\ that
topological closed strings amplitude $F_g(t_i)$ at genus $g$, 
where $t_i$ are the K\"ahler moduli of a Calabi-Yau threefold, computes
the low energy effective theory terms of the Type IIA compactification,
\eqn\clos{\int d^4 x d^4 \theta F_g(t_i) \left( W^2\right)^g,}
where the integral is on the $4d$ $N=2$ superspace, $t_i$
denote a vector multiplet with the K\"ahler expectation values
as the top element.  $W_{\alpha \beta}$ denote the graviphoton
multiplet (with  self-dual
graviphoton field strength as the top component), $\alpha,\beta$
denote symmetric spinor indices and
$$W^2=W_{\alpha \beta}W_{\alpha'\beta'}\epsilon^{\alpha \alpha'}
\epsilon^{\beta\beta'}.$$
In fact it was through this connection where $F_g(t_i)$
were reinterpreted in \gopva\
in terms of spectrum of wrapped M2/D2 branes in the Calabi-Yau threefold.
In particular it was shown that
$$\sum_g F_g(t_i) \lambda^{2g-2}=\sum_{n=1}^\infty 
\sum_{Q\in H_2(M,{\bf Z})}\sum_{s=0}^{\infty} N_{Q,s}
\left(2 {\rm sin}(n\lambda /2)\right)^{2s-2}{e^{-n t_Q} \over n},$$
where $t_Q=\int_Q k$ is the area of the cycle and 
$N_{Q,s}$ denotes the (net) number of $M_2$ brane
bound states of charge $Q$ and $SU(2)_L$ content 
$\left[2(0)+({1\over 2})\right]^{\otimes s}$
(for more detail see \gopva ). This was obtained
by computing the effective one-loop Schwinger-type correction 
 to the terms of the form $R_+^2F_+^{2g-2}$, with
D$2$ brane bound states going around the loop
\ref\nartay{I. Antoniadis, E. Gava, K. S. Narain and T. R. Taylor,
``$N=2$ Type II - Heterotic Duality and Higher Derivative
$F$-Terms,'' Nucl. Phys. B455 (1995) 109; {\tt hep-th/9507115}.}.  The sum
over $n$ above arises because every
D$2$ brane can bind exactly once to an arbitrary number of
D$0$ branes, $i.e.$ every M$2$ bound state
can have arbitrary momentum around the circle. 
In other words the sum over arbitrary
number of D$0$ branes gives rise to a delta function, 
which effectively replaces the Schwinger time integral
by a discrete sum represented by $n$ above.  The
factor of $\left( 2\sin(n\lambda/2)\right)^{2s-2}$ arises from
a $\left( 2\sin(n\lambda/2)\right)^{2s}$ having to do with 
the extra contribution
of a states of spin content $\left[2(0)+({1\over 2})\right]^{\otimes s}$
 running around the loop
in the Schwinger computation, as compared
to a spin 0 which would give $\left(2\sin(n\lambda/2)\right)^{-2}$.

We would like to repeat an analogous scenario for
reinterpretation of topological $A$-model  with D-branes
which include a supersymmetric 3-cycle in the internal Calabi-Yau
threefold as its worldvolume.  There are various cases one can consider.
We will consider in particular the Type IIA compactification
on a Calabi-Yau, with one additional D$4$ brane
wrapped around a supersymmetric 3-cycle $S$
and filling an $R^2\subset R^4$ in the uncompactified spacetime.
Suppose the first Betti number of $S$ is $r=b_1(S)$.  Then
on $R^2$ subspace of $R^4$ live $r$ (2,2) supersymmetric
 chiral superfields
$\Sigma^i$
corresponding to the scalar moduli of $S$ in the
Calabi-Yau threefold \ref\syz{A. Strominger, S.-T. Yau
and E. Zaslow, ``Mirror Symmetry is T Duality,''
Nucl. Phys. B479 (1996) 243; {\tt hep-th/9606040}.}.
The top component of this chiral field is a complex field
whose phase is related to the expectation value of the Wilson line 
of the gauge field $A$ on the D4-brane around
the corresponding 1-cycle of $S$.  Moreover $\Sigma^i$ can
be viewed as a $(2,2)$ vector multiplet on $R^2$.  The $U(1)$
gauge field on $R^2$ corresponds to the magnetic 2-form $B$
field on the D$4$ brane $dB=*dA$ and taking the component
of $B$ along the corresponding cycle in $S$, to yield a gauge field
on $R^2$. One could generalize this by considering $M$ copies
of the D$4$ brane.  We will get in this case $M$ copies of the
$U(1)$ gauge field and so the fields $\Sigma_i^{\alpha}$ will be
naturally in the adjoint of $U(1)^M$.  The permutation
groups ${\cal S}_M$ which is the symmetry of D4 branes acts on this set of
fields to permute the $\Sigma_i$.  Giving vev to 
$\langle\Sigma_i^{\alpha}\rangle =\theta_i^\alpha$ allows us to think of each
$i$-direction a diagonal $U(M)$ matrix of holonomy given
by diagonal elements ${\rm exp}(i\theta_i^{\alpha})$.  Let us
denote this $U(M)$ matrix by $V_i$.

Now we are ready to state what physical amplitude the topological string
computes in the presence of D-branes.  The topological strings
in this case computes
\eqn\opto{\int d^4x d^4\theta \delta^2(x)\delta^2 (\theta)
F_{g,h}(V_i,t)\left( W^2\right)^g \left(W\cdot v\right)^{h-1},}
where $W\cdot v=W_{\alpha \beta}v^{\mu \nu}\gamma^{\mu \nu}_{\alpha
\beta}$, and $v^{\mu \nu}$ denotes the vector orthogonal
to the noncompact worldvolume of D$4$ brane, and $\gamma^{\mu \nu}$
are the usual gamma matrices.  The delta function
above localizes the contribution to the superspace
defined by the noncompact part of the D$4$ brane. Here
\eqn\togw{F_{g,h}(V_i,t)=\sum_{n^i_j}F_{g,n^i_\alpha}(t)
\prod_{\alpha=1}^{h} {\rm tr} \otimes_{i=1}^{b_1(S)}
V_i^{n^i_\alpha},}
and $F_{g,n^i_\alpha}(t)$ denotes the topological string amplitude
at genus $g$ with $h$ holes, labeled by $\alpha=1,\cdots,h$ and where
on each hole $\alpha$ the circle on
the Riemann surface is mapped to the boundary of $S$
 characterized by the $H_1(S)$ class $n^i_\alpha$.
 The trace factors above are just the usual Chan-Paton factors.
 The derivation of \opto\ is similar to that for
 the closed string case and can be done most conveniently in the Berkovits
 formalism \ref\berk{
N. Berkovits, {\sl private communication}.}, similar
to what was done in the closed string case
 for the Calabi-Yau topological
 amplitudes in \ref\berkv{N. Berkovits and C. Vafa, 
``$N=4$ Topological Strings,'' Nucl. Phys. B433 (1995) 123; 
{\tt hep-th/9407190}.}.

As in the closed string case, we would like to connect
  \opto\ 
 with contributions due to wrapped D2 branes.
 The main additional ingredient in this case is
 that the D2 brane can end on the D4 brane $S$.  This
 will give a state magnetically charged under the $U(1)^M$
 living on the D4 brane.  One term included in \opto\
 after doing the superspace integral is a term of the form
 $R F^{2g-2+h}$ where $F$ denotes the expectation
 value of the $4d$ graviphoton field strength restricted
 to the uncompactified worldvolume of the D$4$ brane.  If we
 give a vev to the graviphoton, 
$\langle F\rangle=\lambda$, this would compute correction
 to $\int R $ as a function of 
 $$F(t,V_i)=\sum_{g,h} F_{g,h}(t,V_i) \lambda^{2g-2+h}.$$
 This is the summed up version of the topological
 string amplitudes over all genera and holes, where
 the role of the string coupling constant $\lambda$
is played by the vev of $F$.

We thus compute the contribution of magnetically
charged D4 branes ending on $S$ to $\int R$ in the presence
of the background $F$.  Each such
particle will transform according to some
representation ${\cal R}$ for $\otimes_i
U_i(1)^M/{\cal S}_M$, where
$i$ runs from $i=1,\cdots,b_1(S)$.  We in principle
do not know if they form representation of $U(M)$ (for each
element of $b_1(S)$)\foot{ There is a priori no reason
why $M$ coincident branes give rise to a {\it magnetic} $U(M)$ gauge
theory.}, but nevertheless we can assign
them to representatitions of $U(M)$ if we allow negative
multiplicity.  This is because any ${\cal S}_M$ symmetric 
spectrum for $U(1)^M$ can be written as combination of weights
appearing in various representations of $U(M)$.  From this point
on, we will therefore take ${\cal R}$ to be a representation
of $U(M)$ (for each $b_1(S)$) and allow negative multiplicities.
In addition
every such state is characterized by its bulk D$2$ brane
charge $Q\in H_2(M,S)$, $i.e$. a 2-cycle in the Calabi-Yau threefold
ending on $S$.
Every such field
will be represented by some spin $s$ field in 2 dimensions,
where $2s$ is a positive or negative integer.
To determine $s$, it
 is most convenient to view it from the M-theory perspective;
In the strong
coupling limit this geometry gets related to M-theory on the Calabi-Yau
threefold, with M5 brane filling $S\times R^3$.  The magnetic
charged state correspond to particles in 3 dimensions  with
$M2$ branes ending on the M5 brane.  The little group
of massive particles in three dimensions is $SO(2)$
and so the particle carries a $3d$ spin $s$. Upon
reduction to $2d$, this particle is realized
by a field with spin $s$.  Moreover each such
particle can be bound to an arbitrary number of D0
branes.  This is clear also from the M-theory perspective
as each particle carries an arbitrary momentum as we
go down from 3 dimensions to 2 on a circle.
The computation then is as in the closed string case,
where we effectively get the Schwinger computation for a scalar
field (with the supersymmetry being responsible for generating
$\int R$), and 
$$F(t,V_i)=i\sum_{n=-\infty}^\infty \sum_{{\cal R}, Q, s}
\int_{0^+}^{\infty} {d\tau\over \tau} 
 {N_{{\cal R},Q,s} \over 
2{\rm sin}(\tau \lambda /2)}e^{is\lambda \tau}
{\rm Tr}~ e^{(-m_{R,Q}+2\pi i n )\tau}.$$
Here $N_{{\cal R},Q,s} $ denotes the net number of magnetically
charged states with charges given by ${\cal R}, Q$ and spin 
$s$. The parameter $\tau$ is the Schwinger time, the sum over $n$ 
is the sum over the D$0$ brane bound states, and
$m_{R,Q}+2\pi i n$ denotes the BPS mass of the wrapped D2 brane,
which is given by
$${\rm Tr} ~ e^{-m_{R,Q}+ 2\pi i n}= e^{-t_Q +2 \pi i n}
\Tr_{\cal R}\prod_{i=1}^{b_1(S)} V_i,$$
where $t_Q =\int_Q k$.
To see how the above expression arises, note that for one D4 brane this follows
from the fact that $\int_Q k$ is just the bulk
contribution to the BPS formula and 
$\Tr_{\cal R}\prod_{i=1}^{b_1(S)} V_i$ arises from the fact that
giving
vev to the $U(1)$ fields for each one gives a 
BPS mass $q\theta, $ where $q$ is the charge
under $U(1)$ and $\theta $ denotes the vev of a the scalar in $U(1)$ multiplet.
Doing the sum over $n$ in the above gives a delta function 
$\sum_{n=-\infty}^\infty
\delta (\tau-n)$, which converts the $\tau $ integral into
a sum, and we obtain
\eqn\finaf{F(t,V_i)=i\sum_{n=1}^\infty  \sum_{{\cal R}, Q, s}
 {N_{{\cal R},Q,s} \over 2n{\rm sin}(n\lambda /2)}
e^{n (-t_Q + is\lambda)}{\rm Tr}_{\cal R}\prod_{i=1}^{b_1(S)} V_i^n.}
Note that to compare it with \togw\ one has to expand the
trace from representation $R$ in terms of fundamental representation
of $U(M)$.  Note that the above expression has strong
integrality predictions which would be interesting to
verify.

Note that for the special case of $g=0,h=1$, i.e. the disc
amplitude \opto\ computes theta terms in gauge theory.  Namely
for each diagonal element of $ V_i$, denoted by $\exp(i \theta^{\alpha}_i)$
the term
$${\partial F_{0,1} \over\partial \theta^{\alpha}_i}(t,V_i),$$
denotes the correction to the theta term $\int F^{\alpha}_i$ where 
$F^{\alpha}_i$ denotes the field strength for the corresponding $U(1)$
gauge field in $2d$.  From \finaf\ we
can read the prediction for this, which is
given by
$$
\eqalign{& i\sum_{n=1}^\infty
\sum_{v^\alpha_i\in {\cal R}, Q,s} N_{{\cal R},Q,s} ~q_i^{\alpha}{1 \over n} 
e^{-n (t_Q+i v^\alpha_i\theta^\alpha_i)} \cr
& =
-i\sum_{v^\alpha_i\in {\cal R}, Q,s} N_{{\cal R},Q,s}
~q_i^{\alpha} {\rm log }
(1-e^{-t_Q-iv^\alpha_i\theta^\alpha_i}) \cr
&=-i\sum_{m=-\infty}^\infty
\sum_{v^\alpha_i\in {\cal R}, Q,s} N_{{\cal R},Q,s}
~q_i^{\alpha}{\rm log}
(t_Q+iv\cdot \theta
+2\pi i m)},$$
which is the expected correction to the theta angle in $2d$ $N=2$ gauge theory
from charged matters with BPS masses $t_Q+iv\cdot \theta
+2\pi i m$ and charge $q^{\alpha}_i$ (see in particular a
similar correction which was studied in \ref\hho{A. Hanany and K. Hori,
``Branes and $N=2$ Theories in Two Dimensions,'' Nucl. Phys. B513 (1998)
119; {\tt hep-th/9707192}.}).  Note that from \finaf\
we can write $F_{0,1}$ in the form\foot{As before we are dropping
terms polynomial in $t$ and $\theta$'s which would have
corresponded to $n=0$
in the above sum.},
\eqn\disa{F_{0,1}(t,V_i)=i\sum_{n=1}^\infty
\sum_{v\in {\cal R},Q,s}
N_{{\cal R},Q,s}
{1 \over n^2}e^{ -n
(t_Q+iv\cdot \theta)}.}

\subsec{Applications to $N=1$, $d=4$ systems}

In the above to interpret the topological string amplitudes
with boundaries, we used a D4 brane system with worldvolume
$R^2\times S$.  Instead we could have used a D6 brane system
with worldvolume $R^4\times S$.  This would only make
sense in the context of non-compact Calabi-Yau manifolds
(otherwise the flux of D6 brane charge has nowhere to go). Then again
the fields $V_i$ correspond to chiral fields in $d=4$.  In this case the
interpretation of the topological amplitudes given in \opto\ gets
modified.  The simplest case to consider turns out to be $h=1$ 
and $g$ arbitrary. In this case the topological string amplitudes
compute
\eqn\oppto{\int d^4x d^4\theta \delta^2 (\theta)
F_{g,1}(V_i,t)\left(W^2\right)^g, }
and in particular $F_{0,1}$, $i.e.$ topological
disc amplitudes computes superpotential terms
for N=1 theories in four dimensions. This has already been noted
in \bcovtwo\ref\douglaw{I. Brunner, M. R. Douglas, A. Lawrence
and C. Romelsberger, ``D-Branes on the Quintic,''
{\tt hep-th/9906200}.}\ and is being further
studied in \ref\kkl{
S. Kachru, S. Katz,
A. Lawrence, and J. McGreevy, ``Open String Instantons and Superpotentials,''
{\sl to appear}.}. Let us call
$F_{0,1}=W(V_i,t)$, the superpotential.
 From the formula \disa\ we thus have a general expression
for the superpotential $W$ in terms of the spectrum of BPS states
namely
\eqn\supep{W(t,V_i)=\sum_{n=1}^\infty
\sum_{v\in {\cal R},Q,s}
N_{{\cal R},Q,s}{1 \over n^2}
e^{-n (t_Q+iv\cdot \theta)} .}
A special simple case of this is when we have a single
brane where $V_i$ can be viewed as a
complex superfield $e^{i\theta_i}$.
Given that our derivation of this term seems to require
2-dimensional concepts, it is natural to ask if we could
also reproduce this from a 4-dimensional viewpoint.  As we
will see this is also possible.  In the case of D6 branes
with worldvolume $R^4\times S$, the magnetically
charge branes are D$4$ branes ending on the D6 brane.  This
will correspond to a domain wall in $R^4$.  The expression 
\supep\ then suggests that we should be able to
relate the superpotential term, to the structure of domain walls
by ``integrating them out.'' However unlike the 2-dimensional
case, we cannot send the domain walls around the loop, so the question
is how would we obtain such an expression by integrating fields
out in the $4d$ case.  

A hint comes from the recent work \ref\hov{
K. Hori and C. Vafa, ``Mirror Symmetry,'' {\sl to appear}.}\ and a similar case studied
in \ref\gvw{S. Gukov, C. Vafa and E. Witten, ``CFT's from Calabi-Yau
Four-folds,'' {\tt hep-th/990670}.}, where it was shown how extra fields
are relevant for reproducing the domain wall structure. 
For each domain wall, we introduce a field $Y_\alpha$ as a chiral
superfield, which characterizes it by shifting by $2\pi i$ as
we go across the domain wall.  Since we can have a priori
an arbitrary number of domain walls, we must thus have
infinitely many vacua, given by shifting the expectation
value of $Y_\alpha $ by $2\pi i n$.  Moreover the
tension for the domain wall should be given by the BPS formula,
\eqn\tbps{W(Y+2\pi i)-W(Y)=2\pi i Z.}
The superpotential satisfying these constraints which was
found in \hov\ in a similar context is given by
$$W=Z_{\alpha} Y_\alpha +{\rm exp}(-Y_{\alpha}).$$
Note that the critical points obeying $d_YW=0$ are given by
$${\rm exp}(-Y_{\alpha})=Z_{\alpha},$$
namely $Y_\alpha=-{\rm log}Z_\alpha +2\pi i n$, and that
the equation \tbps\ is satisfied. If we integrate out the
hidden variable $Y_\alpha$ we obtain by replacing
$Y_{\alpha}$ by its critical value a superpotential term
\eqn\anonam{W=Z_{\alpha}(1-{\rm log} Z_\alpha) .}
In the case at hand for each element $v$ in a
representation $v \in {\cal R}$
of the magnetic charges, and charge $Q$ in the bulk
we have $N_{{\cal R},Q,0} $ net BPS domain walls for each integer
m, with BPS tension
$$Z_{\alpha}=t_Q+iv\cdot \theta +2\pi i m.$$
Plugging this into \anonam\ and summing over all such
states, we obtain the formula \supep .

\subsec{Application to D-brane on ${\cal O}(-1)+{\cal O}(-1)$ over $S^2$}
In this section we will show how the results
of the previous section are in agreement with the above analysis,
and in particular gives an independent derivation for topological
$A$-model in ${\cal O}(-1)+{\cal O}(-1) $ over $S^2$ with the 
$3$ cycle $\CC $ we have
discussed.  In that case the $b_1(\CC )=1$ and so we have only one
chiral field, which gives rise to one holonomy matrix $V$.
It is clear what the magnetically charges states are; they correspond
to the D2 brane wrapping the northern hemisphere and ending on $S$
or the one wrapping the southern hemisphere and ending on $S$.
The first one has $(t_Q,{\cal R})$ given by $(t/2,$ fundamental$)$
and the second has $(t/2,$ anti-fundamental$)$.  They both have
spin 0, as there is no moduli for them.  We thus obtain
from \finaf :
$$F(t,V)=i\sum_{n=1}^\infty 
{{\rm tr }V^n+{\rm tr} V^{-n}\over 2n \ {\rm sin}(n\lambda/2 )}
e^{-nt/2},$$
which agrees with the knot invariant predicted for the unknot,
as discussed in the previous section, with $t=2\pi iN/(k+N)$
and $\lambda =2 \pi/(k+N)$.

\newsec{Suggestions for Future Work}
Here we have mainly concentrated in computing the
expectation value of the Wilson loop for a simple knot,
and have found a striking agreement with the predictions of topological
strings in the Large $N$ limit, anticipated from the large
$N$ Chern-Simons/topological string duality proposed in \gvconjecture .
This was done by independently computing both sides and
checking that they agree. On the topological string we used
D-branes ending on branes to get a prediction for what
the topological string should reduce to.

It would be nice to generalize this for arbitrary knots. There
are two obstacles to overcome.  On the Chern-Simons
side we need to compute 
$$ \langle Z(U,V) \rangle_{S^3} = 
\left\langle {\rm exp}\left[
\sum_{n=1}^\infty \tr {1 \over n} U^n {\rm tr}V^{-n} \right]
\right\rangle_{S^3} .$$
This is already rather difficult to do, even though in principle
it should be possible.  The reason for this is the appearance
of all the powers of ${\rm tr} U^n$.  In particular we
need to know all correlations $\langle{\rm tr} U^{n_1}\cdots{\rm tr} 
U^{n_k}\rangle$. For a general knot, the correlators do not 
decouple, unlike the unknot \productproduct.  
Even though it is in principle possible to compute them, they have not
been computed in the full generality we need.  Nevertheless the structure
of the answer for the $\langle{\rm tr} U\cdots{\rm tr} U\rangle$ dictated by
the Skein relations \wittencs\
are compatible with the general answer expected for
the knot invariants, which follows from the discussion in the previous section,
in particular \finaf .  Note that  we are mapping all the knot
invariants for arbitrary representations, into new integer
invariants $N_{Q,{\cal R},s}$, where two different $Q$'s differ
by an integer (so they can be parametrized by an integer),
$s$ denotes a positive (or zero) spin representation, and ${\cal R}$
is a representation of $U(M)$ for any $M$.  We expect
that for each knot ${\cal R}$ will stabilize for large enough $M$.
What we mean by this is that it will given by representations
with finite number of boxes in the Young tableau (or whose
conjugate has finite number of boxes).  Thus for $M$ large
enough we have `probed' the full content of ${\cal R}$
representation
(for example for the unknot we found that already $M=1$ is sufficient
and the structure for higher $M$'s can be induced from that).
This may be a very useful reformulation of knot invariants, somewhat
analogous to the reformulation of Gromov-Witten invariants, in 
terms of the new invariants defined in \gopva .  In particular the knot
invariants would be given by\foot{In comparing
with the knot invariants, it is natural to do analytic continutation,
as we have seen in the case of the unknot, in the case where
the representation are tensor products of the fundamental
representation of $U(M)$ (as opposed
to tensor products of the anti-fundamental representation).  
This would be equivalent to replacing
for those representations
the corresponding $t_Q\rightarrow -t_Q$, and 
changing the sign of the power of $V$ and putting an overall sign
in front of those terms.}
$$\langle Z(U,V) \rangle_{S^3}=\exp \left[ i\sum_{n=1}^\infty
 \sum_{{\cal R}, Q, s} {N_{{\cal R},Q,s} \over 2 n{\rm sin}(n\lambda/2)}
e^{n (-t_Q+ is\lambda)} {\rm Tr}_{\cal R} V^n \right]$$

For understanding this new formulation of knot invariants,
we also have to construct a Largrangian submanifold for an
arbitrary knot, on the resolution of the conifold,
generalizing our explicit construction for the unknot.  That there
should be such a canonical Lagrangian submanifold for each knot
is natural.  This
is because we already have identified, for an arbitrary knot, the Lagrangian
submanifold on the $T^*S^3$ side, and small resolution does not change
the geometry of the Lagrangian submanifold at infinity.  So with some
deformation near the origin we should be able to obtain the Lagrangian
submanifold after the conifold singularity is blown up.  Then we are 
predicting that the topological string amplitudes, whose answer must
have the structure \finaf , compute the knot invariants.  This
would be a very important subject to develop, not only for
a deeper understanding of knot invariants, but also for a better
understanding of topological strings with boundaries.

Also we have seen in this paper how we can compute superpotential
terms on the Type IIA compactifications on Calabi-Yau threefold
in the presence of D6 brane, at least in some simple cases.
In a more general case, doing the computation on the mirror
should be simpler \ref\vex{C. Vafa, ``Extending Mirror Conjecture
to Calabi-Yau with Bundles,'' {\tt hep-th/9804131}.}.  
Some examples of this have been recently
studied in \kkl .  This may well lead to a method
for geometric engineering of $N=1$ and its solution in terms
of the type IIB mirror.  Namely, we start with the usual
geometric engineering of $N=2$, introduce additional
D$6$ branes to break the $N=2$ to $N=1$ (effectively
giving mass terms to the adjoint fields) and then using
the type IIB mirror to compute the superpotential terms
generated, very much the way prepotential
for $N=2$ theories were computed using mirror symmetry.
This would be very exciting to develop further.

\bigskip

\noindent
{\bf Acknowledgements:}

We are grateful to N. Berkovits, R. Gopakumar, K. Hori, 
S. Sinha, C. Taubes and T. Taylor 
for valuable discussions.
H.O.\ would like to thank the theory group of Harvard University, 
where this work was initiated and completed.

The research of H.O.\ was supported in part by 
NSF grant PHY-95-14797, DOE grant DE-AC03-76SF00098,
and the Caltech Discovery Fund. 
The research of C.V.\ is supported by NSF Grant No. PHY-9218167. 

\vglue 2cm
\centerline{\bf Note Added}
After the completion of this work, beautiful
computations were done
\ref\lm{J. Labastida and M. Mari\~no, ``Polynomial
Invariants for Torus Knots and Topological Strings,'' to appear.}\
for checking the predictions made in this paper for the case
of torus knots and finding impressive agreement with what
was anticipated.
We would like to thank J. Labastida and M. Mari\~no,
who informed us of their computation prior to publication
which prompted us to correct an error we had made in the Schwinger
 computation in an earlier version of this paper.

\vfill\eject

\listrefs

\end